\documentclass{aa}
\usepackage[]{graphics}
\usepackage[]{psfig}
\usepackage[]{epsfig}
\begin{document}

\title{Ultra Compact Objects in the Fornax Cluster of Galaxies: Globular
  clusters or dwarf galaxies?}
\author {Steffen Mieske \inst{1,2} \and Michael Hilker \inst{1} 
\and Leopoldo Infante \inst{2}
}
\offprints {S.~Mieske}
\mail{smieske@astro.uni-bonn.de}
\institute{
Sternwarte der Universit\"at Bonn, Auf dem H\"ugel 71, 53121 Bonn, Germany
\and 
Departamento de Astronom\'\i a y Astrof\'\i sica, P.~Universidad Cat\'olica,
Casilla 306, Santiago 22, Chile
}
\date{Received 30 October 2001 / Accepted 19 December 2001}
\titlerunning{Ultra Compact Objects in Fornax - GCs or dEs?}
\authorrunning{S.~Mieske et al.}

\abstract{The relation between the Ultra Compact Objects (hereafter UCOs) recently discovered in the Fornax cluster
(Drinkwater et al. \cite{Drinkw00a}, Hilker et al. \cite{Hilker99}) and the brightest
globular clusters associated with the central galaxy NGC 1399 has been investigated. The
question was adressed whether the UCOs constitute a
distinct population of objects not linked to globular clusters or
whether there is a smooth transition between both populations.\\
Therefore, a spectroscopic survey on compact objects in the central region of the
Fornax cluster was carried out  with the 2.5m du Pont telescope
(LCO). UCOs and the bright NGC 1399 globular clusters with similar
brightness were inspected. 12 GCs from the bright end of the globular
cluster luminosity function have been identified as Fornax
members. Eight are new members, four were known as members from
before. Their magnitude distribution supports a smooth transition
between the faint UCOs and the bright globular clusters. There is no
evidence for a  magnitude gap between both populations. However, the brightest
UCO clearly stands out, it is too bright and too large to be accounted for by
globular clusters. For the only UCO included in our survey, a
relatively high metallicity of $[\frac{Fe}{H}]\simeq -0.5$ dex is measured.\\
\keywords{galaxies: clusters: individual: Fornax cluster -- galaxies:
dwarf -- galaxies: fundamental parameters -- galaxies: luminosity
function -- globular clusters: luminosity function}}
\maketitle

\section{Introduction}

\subsection{Magnitude - surface brightness relation of early type dwarf galaxies}

The faint end of the galaxy luminosity function is mainly populated by
dwarf elliptical galaxies (dEs) and dwarf spheroidals (dSphs,
the faintest dEs in the Local Group). These galaxies
 are the most numerous type
of galaxies in the nearby universe, having absolute magnitudes fainter
than $M_{V}$ $\simeq$ $-$17 mag. They follow a tight magnitude-surface
brightness relation in the sense that central surface brightness
increases with increasing luminosity (Ferguson \& Sandage \cite{FergSan88},
\cite{FergSan89}). The validity of this relation has been a subject of lively
debate over the last decade. A number of  authors 
have argued against the existence
of a magnitude-surface brightness relation for dEs (Davies et
al. \cite{Davies88}, Phillipps et al. \cite{Philli88}, Irwin et al. \cite{Irwin90}) and questioned the
cluster membership assignement to dEs based on morphology.\\
Recently, spectroscopic membership confirmation has
shed light into this matter. Drinkwater et al. (\cite{Drinkw01a}) do confirm the
brightness - surface brightness relation for Fornax dwarfs based on 
the data of their Fornax Cluster 
Spectroscopic Survey (FCSS). They obtain spectra for all
objects in the central 2 degrees of the Fornax cluster down to a
limiting magnitude of $M_V\sim-12.5$ mag and find that the
magnitude surface brightness relation for dEs is well defined. Under
this scheme,
galaxies whose total luminosity and/or surface brightness lie
significantly outside this relation are hard to classify. 
Two of the few examples for such peculiar
objects that have been known for a long time are the high surface
brightness compact dwarf elliptical (cdE) M32 and the very extended
low surface brightness spiral Malin 1 (Bothun et al. \cite{Bothun87}, Impey et
al. \cite{Impey88}, Bothun et al. \cite{Bothun91}).\\

\subsection{New Ultra Compact Objects}
\begin{figure}[h!]
\vspace{-0.5cm}
\epsfig{figure=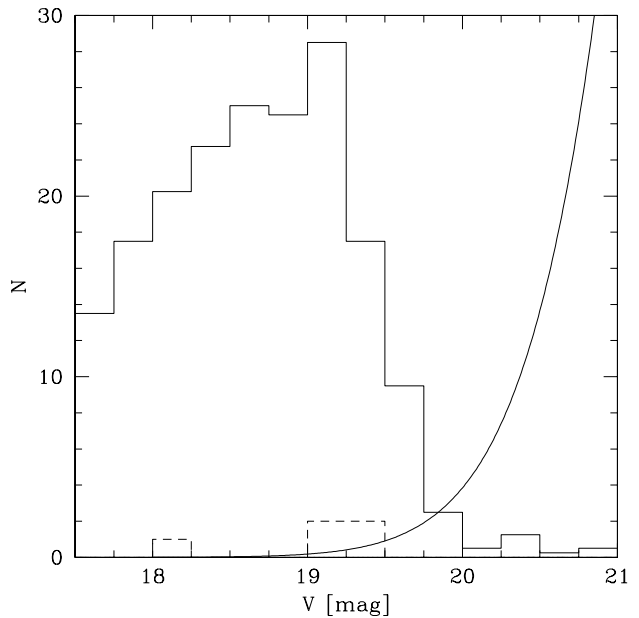,width=8.6cm,height=8.6cm}
\caption{\label{introlf}{\it Solid line}: GCLF of NGC 1399, represented by a Gaussian with
$V_{\rm to}=23.9$ mag and $\sigma=1.2$ mag. Both values taken from
Kohle et al. (\cite{Kohle96}). {\it Solid histogram}: LF of all observed
sources in the FCSS in the central Fornax cluster, divided by four to
fit into the plot (Drinkwater et al. \cite{Drinkw01c}, $V$ magnitudes
 from Hilker et al. \cite{Hilker02}, in prep.). It is
shown to illustrate the FCSS's completeness limit (90\% at about $V=19$
mag). {\it Short dashed histogram}: LF of the UCOs.}
\end{figure}

Most recently, in the course of the FCSS, Drinkwater et al. (\cite{Drinkw00a})
detected five ultra compact objects (UCOs, UCDs in their papers) within
30$'$ projected distance from the Fornax cluster's central galaxy, NGC 1399\footnote{Two of them
had already been detected by Hilker et al. (\cite{Hilker99})}. Although as
bright as average size dEs, they are by far more compact.
Four of the
five UCOs have absolute magnitudes of
about $M_V = -12$ mag, one is significantly brighter with $M_V = -13.3$ mag. On HST-STIS images, the four fainter UCOs have King profile effective radii between 10 and 17 pc, while the brightest one has about 50 pc (Drinkwater et al. \cite{Drinkw01b}). In Table \ref{ucocor}, the properties
of the UCOs are summarized. 
All of them are significantly brighter than the
brightest galactic globular cluster ($\omega$ Centauri has $M_V=-10.2$
mag) but significantly fainter than M~32 ($M_V=-16$ mag). Neither
Hilker et al. (\cite{Hilker99}), Drinkwater et al. (\cite{Drinkw00a}, \cite{Drinkw01b}) nor Phillipps et
al. (\cite{Philli01}) could draw definite conclusions
about the nature of the UCOs.\\
\begin{table*}
\caption{\label{ucocor} Properties of the five
UCOs. Adopted distance modulus to Fornax 31.3 mag. Magnitudes are from a
photometric wide field survey of the central Fornax cluster (Hilker et
al. \cite{Hilker02}, in prep.). $^*$ UCO~3 and 
4 were detected first by Hilker et al. (\cite{Hilker99}).}
\begin{flushleft}
\begin{center}
\begin{tabular}{llllllll}
\hline\noalign{\smallskip}
Name & $\alpha$ (2000) & $\delta$ (2000) &$V$ [mag] & $M_V$ [mag]&$(V-I)$
[mag]&$r_{\rm eff}$ [pc]&$d$(NGC 1399) [$'$]\\\hline\hline
UCO~1 & 03:37:03.30 & -35:38:04.6 & 19.31 & -11.99 & 1.17&12&20.74\\
UCO~2 & 03:38:06.33 & -35:28:58.8 &19.23 & -12.07 & 1.14&15&5.10\\
UCO~3$^*$ & 03:38:54.10 & -35:33:33.6 &18.06 & -13.24 & 1.20&50&8.26\\
UCO~4$^*$ & 03:39:35.95 & -35:28:24.5 &19.12 & -12.18 & 1.14&17&13.64\\
UCO~5 & 03:39:52.58 & -35:04:24.1 &19.50 & -11.80 & 1.04&10&28.26\\
\hline
\end{tabular}
\end{center}
\end{flushleft}
\end{table*}One possibility is that UCOs are bright globular clusters. NGC 1399 has a
very rich globular cluster system with about 6000 GCs within 10$'$ (about 40
kpc at Fornax distance) from its center (Kohle et al. \cite{Kohle96}, Forbes et
al. \cite{Forbes98}) which may contain GCs as bright as the UCOs.
Another possibility is that UCOs are
the nuclei of stripped dwarf galaxies. Threshing dE,Ns in the cluster potential has been shown to work
(Bekki et al. \cite{Bekki01}). Lotz et al. (\cite{Lotz01}) find that the luminosity
function of 27 Fornax and Virgo dwarf nuclei peaks at $V=21.7$ mag
($M_{V}=-9.6$ mag) with a dispersion of $\sigma=1.2$ mag, whereas the Globular Cluster
Luminosity Function (GCLF) peaks at about $M_V=-7.4$ mag. So, nuclei are on average
more than 2 mag brighter than GCs and might mix up with the bright
tail of the GCLF. Or do the UCOs represent a new group of ultra
compact dwarf galaxies, extreme cases of M~32? No discriminating
statement could be made until now.\\
\noindent
It would be very interesting to know whether the UCOs had an origin different from the population of 
globular clusters of NGC 1399.\\
The UCOs have magnitudes roughly equal to the completeness magnitude
limit of the FCSS ($V=19$ mag or $M_V=-11.7$ mag). This is about 4.5
magnitudes brighter than the turnover of the globular cluster luminosity
function (Kohle et al. \cite{Kohle96}). In Fig.~\ref{introlf}, the three
relevant luminosity functions are shown in one plot: the LF of all
observed sources in Drinkwater et al.'s FCSS (to indicate
its completeness limit); the LF of the UCOs;
and the bright end of NGC 1399's GCLF, respresented by a Gaussian with
$V_{\rm to}=23.9$ mag and $\sigma=1.2$ mag, as taken from Kohle et
al.'s GCLF. A total number of 8100 GCs was adopted, which is the number
contained within 20$'$ from NGC 1399 (see Sect.~\ref{totalnum}).
\\
\\
\subsection{Aim of this paper}
As one can see in Fig.~\ref{introlf}, 
the magnitude range between the UCOs and the
bright globular clusters must be probed more
thoroughly, both  to know whether
UCOs extend to fainter magnitudes and to determine the bright end of
the GCLF. A gap in magnitude space between both populations would imply that
the UCOs are very compact dEs or nuclei of
stripped dwarfs rather than globular clusters.\\
In this paper, we describe and analyse a survey of compact objects in
the central region of the Fornax cluster of galaxies 
that closes the magnitude gap
between the FCSS and the globular cluster regime.\\
In Sect.~\ref{Selcand} we describe the selection of candidates for the
survey. In Sect.~\ref{obsdatared} the observations and data reduction
are described. Sect.~\ref{measureradvel} shows the radial velocity
measurement. In Sect.~\ref{analysis} the results are analyzed, the
question whether or not the UCOs are a distinct population is
discussed and metallicities of a number of Fornax dE,Ns are
measured. These results are discussed in
Sect.~\ref{discussion}. Finally, in Sect.~\ref{summary} a summary and
conclusions are presented.\\
\section{Selection of candidates}
\label{Selcand}
To choose candidates for our spectroscopic survey, we analyzed
existing wide field images of the Fornax cluster. Those had been
obtained over four nights in December of 1999 with the 2.5m Du Pont
Telescope at Las Campanas Observatory, Chile. Their field of view was
25$'$ $\times$ 25$'$; 14 fields were observed. These fields map a
circular region of 2 degrees diameter around the central giant
elliptical galaxy, NGC 1399. The seeing ranged from 1.5$''$ to 2$''$,
which corresponds to a spatial resolution of $\approx$ 120 - 160 pc at
the distance of the Fornax cluster, assumed to be 18 Mpc (Kohle et
al. 1996). The pixel scale was 0.774$''$/pixel. Images in $V$ and $I$
bandpasses were obtained for all fields. One of the central fields was
also observed in $B$. Details of the photometry and of the
calibrations will be presented in an accompanying paper (Hilker et al. \cite{Hilker02},
in prep). All object magnitudes mentioned in this work are taken from that paper.\\
\subsection{Unresolved objects}
\begin{figure} 
\psfig{figure=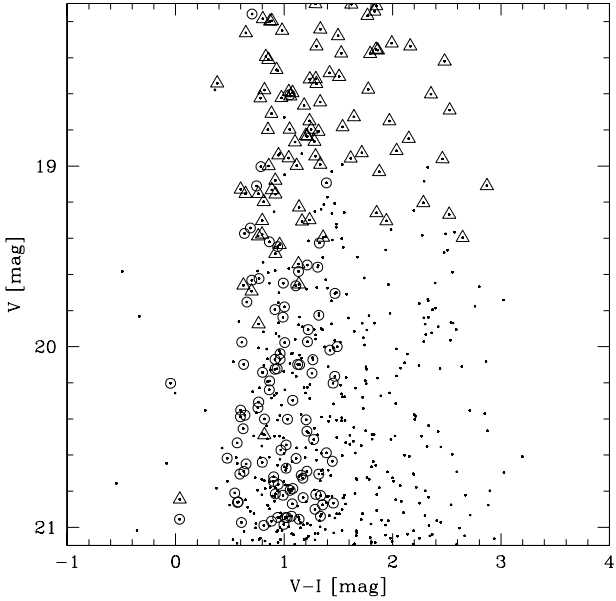,width=8.6cm}
\caption{\label{cmdseldw2} Colour magnitude diagram of one of the
selected fields, restricted to $18.1<V<21.1$. Points are all detected
sources; open circles indicate sources satisfying our selection
criteria for unresolved objects; open triangles are objects included
in the FCSS (Drinkwater et al. 2000, private communication).}
\end{figure}

In turn, we describe the {\it compact} objects selection
criteria for spectroscopy. These criteria are based on object brightness,
colour and morphology.\\
The faint magnitude limit was set by analyzing the GCLF of NGC 1399, as determined by Kohle
et al. (\cite{Kohle96}). They adopt a Gaussian distribution for the GCLF and
estimate a turn over magnitude of $V_{\rm to}$=23.9 mag ($M_V=-7.4$ mag) and a dispersion of
$\sigma$=1.2 mag.  Since one of our goals was to include bright
globular clusters (GCs), we defined $V_{\rm faint}$=21 mag as the faint
magnitude limit for our observations, which is 2.4 $\sigma$ brighter
than the turn over. This means that about 0.8\% of all globular
clusters around NGC 1399 would be accessible to our survey. Adopting  6000
as the total number (Kohle et al. \cite{Kohle96}), then about 50 GCs
would enter in our survey. The bright magnitude limit is given by the FCSS, which covers basically
all objects down to $V$ $\approx$ 19 mag. \\
Apart from pure brightness limits, other restrictions had to be
made. Four of the five UCOs are unresolved on our CCD images due to
their small diameter of about 0.2$''$ or 15 pc at Fornax distance (Drinkwater et al. \cite{Drinkw01b}). 
We did therefore not expect compact Fornax members in the magnitude
regime fainter than the UCOs to be resolved on our images. Thus, unresolved sources were interesting
in the first place. 
A source on our images was defined as unresolved, when SExtractors
star-classifier value (0 for a ``perfect'' galaxy, 1 for a ``perfect'' star, see Bertin \& Arnouts \cite{Bertin96}) was larger than 0.45 in both
$V$ and $I$. It was defined as resolved when this was not the case.\\
Of the unresolved objects in the mentioned magnitude regime, only
those which had $(V-I)<1.5$ mag were accepted as candidates. The value
1.5 comes from the following consideration: we wanted to include all
possible kinds of galactic nuclei or bright globular cluster type
objects, but exclude background galaxies at high redshift or very red
giants in the foreground. According to Worthey's (\cite{Worthe94}) stellar
population models, for a very old population of 15 Gyrs with
$[\frac{Fe}{H}]=0$ one gets $(V-I)$=1.33 mag. Taking into account
model uncertainties, photometric errors - which were smaller than
0.05 mag - and reddening by dust (Schlegel et al. (\cite{Schleg98}) found
E(B-V) = 0.013 towards Fornax), $(V-I)$=1.5 mag was chosen as the red
colour limit. 
To avoid missing very young star forming nuclei, no blue limit was applied.\\
Fig.~\ref{cmdseldw2} shows a colour magnitude diagram for one of the
selected fields, with the selected candidates for the
survey and the objects marked.\\
The only UCO~included in our survey is UCO~2. The other ones lie outside the
fields we covered (see Sect.~\ref{analysis}).\\

\subsection{Resolved objects}
The only UCO~resolved on our images (UCO~3) is about 1 mag brighter than the
magnitude completeness limit of the FCSS. Yet, it cannot be ruled out a
priori that there are compact, but resolved Fornax members fainter than the completeness limit of the FCSS. Therefore
we selected in addition to the unresolved objects the smallest resolved sources in the same magnitude colour range
like the unresolved objects. They were given lower priorities in the mask creation process than
the unresolved ones.\\
Apart from compact objects, we included the dE,Ns cataloged in Fergusons's Fornax Cluster Catalog (FCC) (Ferguson
\cite{Fergus89}) that were bright enough and in the field we wanted to
observe. This made it possible to compare the spectral properties of
these nucleus-dominated spectra with those of the UCOs.\\
\section{Observations and data reduction}
\label{obsdatared}
The observations were performed in the three nights of 2000/12/30 to 2001/01/01 at the 2.5m Du Pont at Las Campanas. The
instrument was the Wide Field CCD (WFCCD) camera, which reimages a
25$'$ field onto a Tek$\#$5-Detector of 2048x2048 pixel with a pixel
scale of 0.774$''$/ pixel. Multi-slit masks and the H \& K grism,
which has a good transmission between 3500 and 6300 {\AA}  and a
resolution of about 1.3 {\AA}  per pixel, were used. As the slit width
in our observation was $\approx$ 1.5$''$ (= 2 pixel), the effective
resolution was in the order of 2.5 - 3 \AA, corresponding to 200 km/s
at 4000 \AA. Four fields were observed, with integration times of
roughly 3 hrs each. In total we obtained spectra of 160 objects (40
per field), of which about 100 were unresolved on our images.\\
All the data reduction was performed with the IRAF-packages IMRED and TWODSPEC. After bias subtraction, the cosmic rays were removed. The
combined dome-flats were normalized to unity and the sky-flats were divided by the
normalized dome-flats. The remaining illumination change along the
slit was measured on these sky-flats. The single object exposures were then
flat-field divided and corrected for the remaining illumination
change. The object exposures were corrected for
the tilting caused by the optics of the WFCCD camera and then combined. After these reduction
steps, the spectra were extracted.\\
\section{Radial velocity measurement}
\label{measureradvel}
Radial velocities were determined by performing Fourier cross correlation
between object and template spectra with the IRAF-task FXCOR.\\
\subsection{Templates}
As templates for the radial velocity measurement we used two of our
standard star spectra (HD 54810 and HD 22879) and one synthetic
spectrum taken from Quintana et al. (\cite{Quinta96}). These three templates
showed the highest correlation peaks when cross correlating them with
the four brightest Fornax dE,Ns included in our survey. 
When measuring the radial velocities with these templates, we accept the
result as correct if the confidence level $R$ (the $r$-ratio of Tonry \&
Davis \cite{Tonry79}, defined as the relative height of the cross-correlation peak
with respect to the neighbouring peaks) of the cross correlation peak
between the template and the object spectrum is larger than 3.5 for all three
templates.
This corresponds to a S/N of about 4 between 4500 and 5000 \AA.\\
\subsection{Results}
\begin{figure*} 
\begin{center}
\psfig{figure=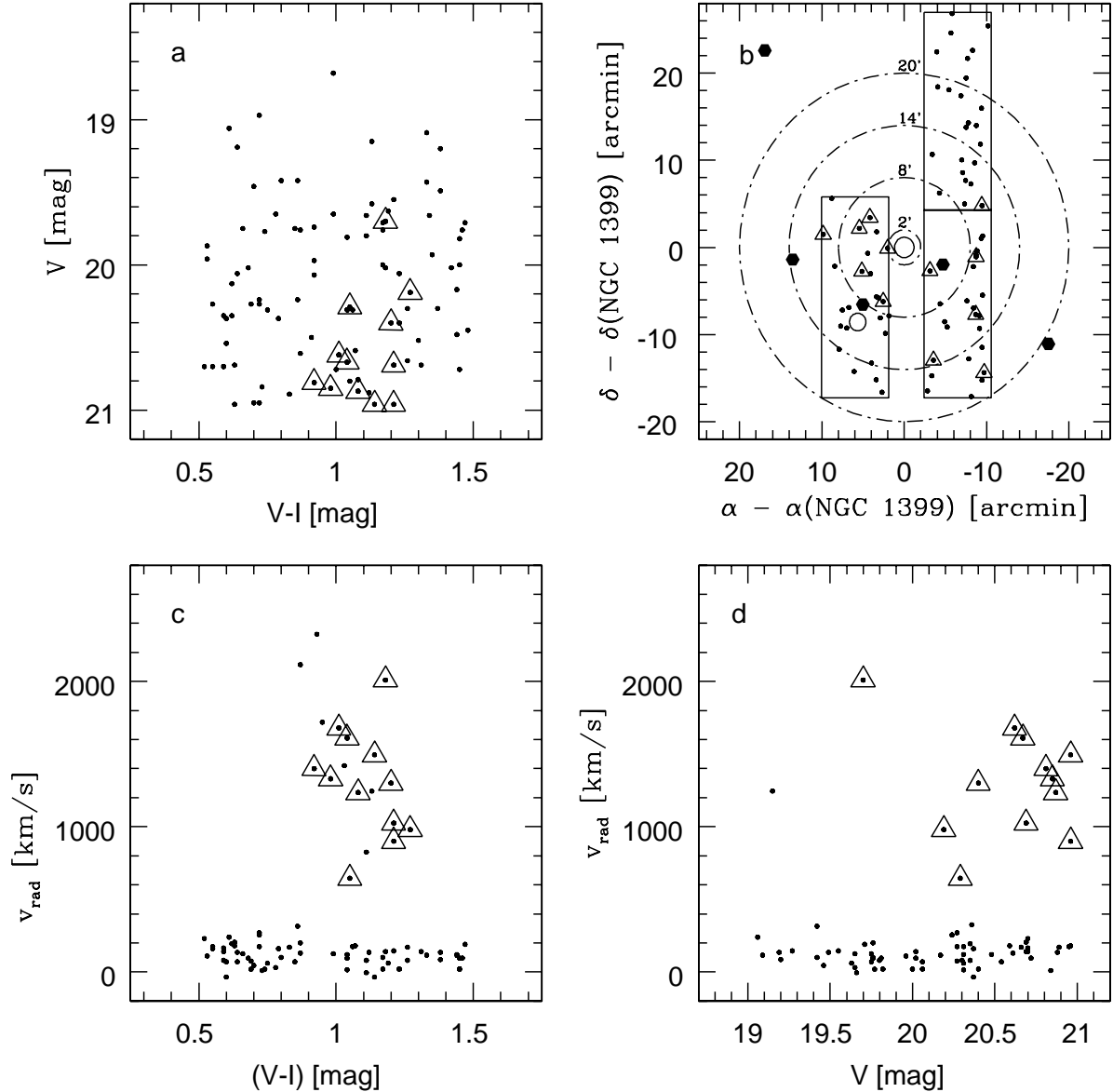,width=17cm}
\end{center}
\vspace{-0.2cm}
\caption[]{Properties of successfully observed sources:
  \label{cmdcvd}\\{\bf a)}: Colour magnitude diagram of the sources
    fainter than $V=18$ mag. {\bf b)}: Position diagram of the point sources
centered around NGC 1399. NGC 1399 and NGC 1404 are indicated as open
circles. Dot-dashed circles indicate distances of 2, 8, 14 and 20$'$ to
1399's center. The field limits are indicated as solid lines. For
comparison, the positions of the UCOs are marked as filled
hexagons. Only UCO~2 in the south west field was observed by us. {\bf c)}:
Colour velocity diagram of all sources, except for background objects. Note that the 6 Fornax detections not
marked with triangles are the 5 dE,Ns and UCO~2. GC candidates are
marked as triangles. {\bf d)}: Magnitude velocity diagram of all sources,
except for background objects and objects fainter than $V=18.7$
mag. Note that the dE,Ns are several magnitudes brighter than $V=18.7$
mag and are therefore not seen. The only Fornax member not marked with
a triangle is UCO~2.\\}
\end{figure*}

In total we obtained 164 spectra in four fields. For 40 of those, the S/N was too low
to reliably determine radial velocities. We successfully determined
radial velocities for 66 foreground stars, 18 Fornax members and 40 background
galaxies. 
Besides five dE,N candidates and UCO~2, 12 unresolved objects
turned out to have a radial velocity between 600 and 2500 km/s, so
they are probable Fornax members. We regard the unresolved
objects as GC candidates (GCCs) from now on, since they are fainter than the
UCOs. They are all within 20$'$ projected distance from NGC 1399. Eight of them have not been
measured spectroscopically before, they are newly discovered members of the Fornax
cluster. Two of the remaining four objects were known to be cluster
members from Hilker (NTT 1998, private communication); the other two
from Kissler-Patig et al. (\cite{Kissle99}). The five dE,Ns and UCO~2 can be
confirmed as cluster members. The parameters of all Fornax members
included in our survey are given in Table \ref{fornmem}. The names of the
dE,Ns are from the Fornax Cluster Catalogue (FCC) of Ferguson
(\cite{Fergus89}). The names of our GC candidates are ``FCOS~Field Number$-$Object
number''. ``FCOS'' stands for ``Fornax Compact Object Survey''. The field
number is: 1 for the south-east field, 2 for the
south-west field and 4 for the north-west field, as shown in Fig.~\ref{cmdcvd}c). The object
number is taken from the mask-creation file. The radial
velocities, their errors and the confidence level $R$ were computed by
averaging the values given from FXCOR for each of the three templates. 
A list of all foreground and background sources is given in the
Appendix. They are available electronically at http://www.XXX. As the
  highest object number is 105 (FCOS~2-105, a foreground star), one leading
zero is used for originally two-digit numbers and two leading zeroes for
originally one-digit numbers, such that all object numbers consist of three
digits.\\
\begin{table*}
\caption{\label{fornmem}Fornax members, ordered by magnitude. See text for
  explanation of names. $R$ is the confidence level of the cross correlation
  peak averaged over the three templates used. GCC stands for ``Globular
  Cluster Candidate''. In the comments-column, values for the radial velocities measured by other authors are given. The references are $^a$ Drinkwater (\cite{Drinkw00a}), $^{b}$
  Drinkwater and Gregg (\cite{Drinkw98}), $^{c}$ Hilker et al. (\cite{Hilker99}), $^{d}$ Held and Mould (\cite{Held94}), $^{e}$ Drinkwater (2000, private communication), $^{f}$ Hilker (NTT 1998, private communication), $^{g}$ Kissler-Patig et al. (\cite{Kissle99}).}
\begin{tabular}[l]{lllrlllll}
\\\hline
Name&$\alpha$ (2000.0)&$\delta$ (2000.0)&$v_{\rm rad} [km/s]$&$R$&$V$&$(V-I)$&Type&Comment\\\hline\hline
FCC~222&3:39:13.23 & $-$35:22:17.6 & 825 $\pm$ 25 &10.7 &14.91 & 1.11 & dE,N& 850 $\pm$ 50$^{c}$\\
FCC~207&3:38:19.42 & $-$35:07:44.3 & 1420 $\pm$ 20 &20.3 &15.34 & 1.03 & dE,N &1425 $\pm$ 35$^{d}$\\
FCC~211&3:38:21.65 & $-$35:15:35.0 & 2325 $\pm$ 15 &31.0 &15.65 & 1.04 & dE,N &2190 $\pm$ 85$^{e}$\\
FCC~B1241&3:38:16.79 & $-$35:30:27.0 & 2115 $\pm$ 25 &11.9 &16.84 & 0.87 & dE,N &1997 $\pm$ 78$^{b}$\\
FCC~208&3:38:18.88 & $-$35:31:50.8 & 1720 $\pm$ 50 &7.0 &17.18 & 1.06 & dE,N &1694 $\pm$ 84$^{c}$\\
UCO~2&3:38:06.41 & $-$35:28:58.2 & 1245 $\pm$ 15 &18.8 &19.15 & 1.13 &UCO&1312 $\pm$ 57$^a$\\\hline
FCOS~1-021 &3:38:41.96 & $-$35:33:12.9 & 2010 $\pm$ 40&7.7 & 19.70 & 1.18 & GCC &1993 $\pm$ 55$^{f}$\\
FCOS~1-060 &3:39:17.66 & $-$35:25:30.0 & 980 $\pm$ 45&8.2 & 20.19 & 1.27 & GCC &\\
FCOS~1-063 &3:38:56.14 & $-$35:24:49.1 & 645 $\pm$ 45&8.1 & 20.29 & 1.05 & GCC &\\
FCOS~2-073 &3:38:11.98 & $-$35:39:56.9 & 1300 $\pm$ 45&4.6 & 20.40 & 1.20 & GCC &$v_{rad}$=1300 {\bf or} 280\\
FCOS~1-019 &3:38:54.59 & $-$35:29:45.8 & 1680 $\pm$ 35&5.6 & 20.62 & 1.01 & GCC &1730 $\pm$ 80$^f$\\
FCOS~1-058 &3:38:39.30 & $-$35:27:06.4 & 1610 $\pm$ 40&5.8 & 20.67 & 1.04 & GCC &1540 $\pm$ 150$^g$\\
FCOS~2-078 &3:37:41.83 & $-$35:41:22.2 & 1025 $\pm$ 60&4.6 & 20.69 & 1.21 & GCC&\\
FCOS~2-086 &3:37:46.77 & $-$35:34:41.7 & 1400 $\pm$ 50&5.74 & 20.81 & 0.92 & GCC &\\
FCOS~4-049 &3:37:43.09 & $-$35:22:12.9 & 1330 $\pm$ 50&5.0 & 20.85 & 0.98 & GCC &\\
FCOS~2-089 &3:38:14.02 & $-$35:29:43.0 & 1235 $\pm$ 45&8.21 & 20.87 & 1.08 & GCC &\\
FCOS~2-095 &3:37:46.55 & $-$35:28:04.8 & 1495 $\pm$ 45&3.9 & 20.96 & 1.14 & GCC &$R$ $<$ limit, 1186 $\pm$ 150$^g$\\
FCOS~1-064 &3:38:49.77 & $-$35:23:35.6 & 900 $\pm$ 85&3.8 & 20.96 & 1.21 & GCC &$R$ $<$ limit\\\hline\hline
\end{tabular}
\end{table*}For three of the 12 GC candidates, namely FCOS~1-064, 2-095 and
 2-073, the membership
assignment is not definite. For
objects 1-064 and 2-095, the confidence level $R$ is lower than 3.5
for one of the templates, but larger for the other two. Cross
correlation with all three templates yields the same (Fornax-)
velocity within the error range. 2-073 has two cross correlation peaks at 1300
km/s and 280 km/s, both with $R$ larger than 3.5 for all templates. The peak
at 1300 km/s is 20\% more pronounced than the peak at 280 km/s.
Object 2-095 is one of the four objects that were
known as Fornax members from before (Kissler-Patig et al. \cite{Kissle99}). This
indicates that our limits for $R$ may be too strict. We therefore
include all 12 GC candidates in our analysis.\\
\section{Analysis}
\label{analysis}
\begin{figure}[h!]
\begin{center}
\psfig{figure=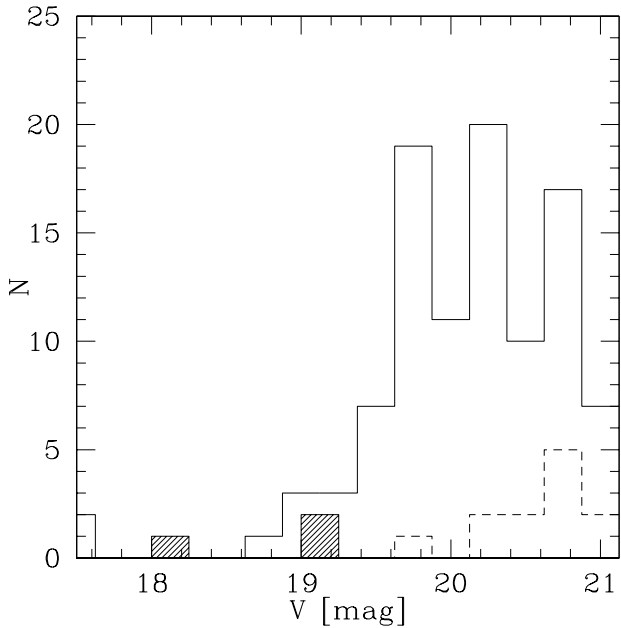,width=8.6cm}
\end{center}
\vspace{-0.2cm}
\caption[]{\label{lfpso} {\it Solid histogram}: Luminosity function of
  all successfully observed unresolved objects. {\it Short dashed histogram}:
  Luminosity function of the GC candidates. {\it Shaded histogram}: Luminosity function of the UCOs within 20$'$ of NGC 1399.}
\end{figure}

In this section a detailed analysis of the objects discovered is presented. From
the total number of GCs associated with NGC 1399 (Dirsch et al. \cite{Dirsch01}, Forbes
et al. \cite{Forbes98}, Kohle et al. \cite{Kohle96}), the form of its GCLF (Kohle et al. \cite{Kohle96}) and
the completeness of our survey we will calculate how many GCs should be
included in our survey and compare that with the number we found. This will
enable us to restrict the form of NGC 1399's GCLF at the bright end. We also determine
whether there is a statistically signifcant gap in magnitude between the
UCOs and our GC candidates. Lick line-indices are
calculated for UCO~2 (see Table~\ref{fornmem})
 and the dE,Ns included in our survey.\\
In Fig.~\ref{cmdcvd}a--d a colour magnitude diagram, a map of the loci of all
successfully observed point sources, a colour velocity and a
magnitude velocity diagram are shown. Note that in field
6, which is more than 1 degree away from the central galaxy NGC 1399,
no GC candidates were found. In Fig.~\ref{lfpso} the luminosity
function for all our detections and for the GC candidates is given, including the 3 UCOs within
20$'$ from NGC 1399. The 2 UCOs outside 20$'$ that were included in
Fig.~\ref{introlf} have been omitted in Fig.~\ref{lfpso}, because our area
coverage outside 20$'$ is basically zero.\\
The mean radial velocity of the GC candidates is 1300 $\pm$ 109 km/s
with a standard deviation of $\sigma$=377 km/s. Richtler et al. (2001, private communication), who obtained
 spectroscopy of about 350 GCs around NGC 1399, get 1447 $\pm$ 16 km/s as
 the mean radial velocity. This is more than 1 $\sigma$ away from our
 result. However, we can rule out a systematic shift in our
 velocity values with respect to the measurements of other authors (see
 Table~\ref{fornmem}). For the ten objects with known radial velocities, the
 median of the difference between literature value and our value is -20 km/s
 $\pm$ 35 km/s. With smaller samples biased to brighter GCs also other
 authors get smaller mean values:
Minniti et al. (\cite{Minnit98}) found for a sample of 18 GCs around NGC
 1399 a mean of 1353 $\pm$ 79 km/s with $\sigma$=338 km/s. Kissler-Patig et al. (\cite{Kissle99}) suggest with a larger sample
 of 74 GCs, that
 the velocity distribution of GCs around 1399 may even be bimodal with two
 peaks at 1200 and 1900 km/s, respectively.\\
The colour magnitude diagram in Fig.~\ref{cmdcvd}a shows that the GC
candidates have colours typical for GCs. Having a mean colour of
$(V-I)=1.11$ with a standard deviation of 0.11, they are only slightly
redder than the average $(V-I)\simeq 1$ for GCs around NGC
1399.\\
\noindent Comparing the luminosity function of all unresolved objects in our
survey with the GC candidates (see Fig.~\ref{lfpso}) shows that the latter
ones are concentrated towards the faint end of our magnitude regime. Only
one of the 12 GC candidates is brighter than $V=20$ mag, although we cover
the whole magnitude range between $19<V<21$ mag ($-10.3<M_V<-12.3$
mag). This drop in frequency towards brighter magnitudes should be expected if all of
the candidates are GCs. We are probing the regime 3 - 5 magnitudes
brighter than the GCLF turnover magnitude ($V_{\rm
  to}\simeq -7.4$ mag) in which GC number counts should decrease towards brighter magnitudes.
This is of course only a qualitative statement. To draw more quantitative
conclusions, two questions need
to be answered: can the number of our GC candidates be explained only
by bright GCs belonging to NGC 1399's GCS? and, is there a significant
magnitude gap between both populations? The next two subsections deal 
with these questions.\\
\subsection{The expected number of globular clusters}
\label{numgc}
How does the number of our GC
candidates compare with the number expected from Kohle et al.'s (\cite{Kohle96}) GCLF? To
find that out, three steps are necessary: First, the total number of
GCs associated with NGC 1399 is calculated (Sect.~\ref{totalnum}). Second, the
completeness of our survey is determined (Sections~\ref{photometricincompleteness} and
\ref{geometricincompleteness}). Then the total number is multiplied with the
completeness and the fraction of GCs brighter than our survey limit (Sect.~\ref{resexpnum}).
As the GC surface density and the
completeness vary with radius, the calculations are performed
separately for the three rings as shown in Fig.~\ref{cmdcvd}b.\\
\subsubsection{Total number of globular clusters associated with NGC 1399}
Our detections reach out as far as almost 20$'$ projected distance. Previous photometric
studies of 1399's GCS, like those of Forbes et al. (\cite{Forbes98}) or Kohle et
al. (\cite{Kohle96}), reached only
10$'$. To cover the zone between 10-20$'$, the latest
results of Dirsch et al. (\cite{Dirsch01}) were used.
Their data was obtained with the CTIO's MOSAIC camera and map the GCS
of NGC 1399 out to
a radius of 20$'$. At 20$'$ they still find a GC surface density 2-3
times higher than in a comparison field 3.5 degrees away from the
cluster center. Dirsch et al. subdivide the region around NGC 1399
into three rings with limits 0$'$--2.5$'$, 2.5$'$--8$'$ and
8$'$--20$'$. In Fig.~\ref{cmdcvd}{\bf b} we adopted the ring
limits equal to theirs (except for the innermost ring). There is a different slope in the radial density profile 
in all three rings. 
It holds in units of arcmin$^{-2}$:
\vspace{-0.5cm}\\
\begin{eqnarray}
r  \epsilon  [0;2.5]'  : \rho(r) = 10^{1.28} \\
r  \epsilon  [2.5;8]' : \rho(r) = 10^{0.98}\cdot r^{-0.77} + 10^{1.42}\cdot r^{-1.46}\\
r  \epsilon  [8;20]' : \rho(r) = 10^{1.46}\cdot r^{-1.41} + 10^{1.27}\cdot r^{-1.50}
\end{eqnarray}
The numbers of Dirsch et al. (\cite{Dirsch01}) are not  corrected for
incompleteness. In order to estimate the fraction of objects missing, 
we integrated their number counts within 10$'$,
yielding 1545 $\pm$ 150 GCs, and divided this by the total number of
GCs within that region derived from previous surveys. We adopted the
mean of the two values of Kohle et al. (\cite{Kohle96}) and Forbes et
al. (\cite{Forbes98}), who get 5940 $\pm$ 570 and 5700 $\pm$ 500,
respectively. The ratio is then 0.27 $\pm$ 0.03.\\
\noindent
With that information at hand, the number of GCs within each
of the three rings, as plotted in Fig.~\ref{cmdcvd}{\bf b}, is calculated by
integrating the surface density law within the rings and
dividing the numbers by 0.27. The GC candidate closest to NGC 1399 is only
2$'$ away, so we integrated Dirsch et al.'s values from 2$'$ to 8$'$
instead of 2.5$'$ to 8$'$.
 The results are: 3800 $\pm$ 460 for ring
1 (2$'$--8$'$), 2340 $\pm$ 280 for ring 2 (8$'$--14$'$) and 1960 $\pm$
230 for ring 3 (14$'$--20$'$).\label{totalnum}\\
\subsubsection{Photometric completeness}
\label{photometricincompleteness}
To determine the incompleteness involved in the source extraction
on the photometric images, artificial
star experiments were performed.
With the IRAF task MKOBJECTS in the ARTDATA package 5000 artificial
stars with Gaussian profiles were put into the $V$ and $I$ images
of the three fields where the new members were found. Their magnitude
ranged between $21.5>V>19$ mag and their colour indices between $2>(V-I)>0$
mag. Poisson
noise was negligible compared to the stars' signal and was
therefore not included. Magnitudes, colours and positions in the frame were randomly
created using a C code.\\
To gain a statistically significant result without altering the
crowding properties in our images, we added 50 times 100 artificial
stars to our observed images. Then SExtractor was run on each of them,
using the same parameters as on the original extraction. The
photometric selection criteria for point sources as defined in
Sect.~\ref{Selcand} were then applied to the SExtractor output catalog. 
\\
The ratio between the number of artificial stars that match the
selection criteria and the number of input artificial stars in that
magnitude-colour-range defines the completeness in the photometric
detection. For the objects in the
magnitude-colour range of the selected candidates ($19.5<V<21$ mag and
$0.4<(V-I)<1.5$ mag, see Fig.~\ref{cmdseldw2}) the result is 0.79 $\pm$ 0.07. There is no significant
magnitude dependence of the photometric completeness in the given
magnitude regime. The differences between input and output magnitudes were
on average smaller than 0.07 mag, with a standard deviation of $\simeq$
0.1 mag. For all object magnitudes mentioned in this paper, the internal error is therefore on
the order of 0.05 mag and for the colours it is about 0.07 mag.\\
\subsubsection{Geometric completeness}
\label{geometricincompleteness}
The geometric completeness is obtained by dividing the surface density
$n_{\rm obs}$ of successfully\footnote{Here we implicitly treat the
objects observed with too low S/N like the ones rejected for geometric
reasons} observed point objects by the surface density $n_{\rm sel}$
of point objects that satisfy our selection criteria and were not
included in the FCSS:\\
Geometric completeness = $n_{\rm obs}/n_{\rm sel}$.\\ 
$n_{\rm sel}$ is distance independent. It was determined by counting
the number $N_{\rm sel}$ of objects satisfying our selection criteria
and not included in the FCSS in the central 20$'$ of the CCD frames of field 1, 2 and
4. The area covered is $\pi * 10^2=314.2$ arcmin$^2$, and therefore the
resulting surface density is $n_{\rm sel}=\frac{N_{\rm
sel}}{314.2'^2}$. The numbers for both $N_{\rm sel}$ and $n_{\rm sel}$
are given in Table \ref{nselect}. The mean of $n_{\rm sel}$ is 0.197 $\pm$
0.022 arcmin$^{-2}$.\\
 \begin{table}
\caption[]{\label{nselect}Total number $N_{\rm sel}$ and surface density
  $n_{\rm sel}$ of objects that satisfy our selection criteria and are in
  the central 20$'$ of the three fields where new members were found.}
\begin{tabular}{ccc}
\hline\small
Field-Nr.& $N_{\rm sel}$ & $n_{\rm sel}$ [arcmin$^{-2}$] \\\hline\hline
1 & 52 & 0.166 \\
2 & 74 & 0.235 \\
4 & 59 & 0.188 \\\hline
& 62 $\pm$ 7& 0.197 $\pm$ 0.022\\
\end{tabular}
\normalsize
\end{table}\begin{table}
\caption[]{\label{nobserv}Surface density $n_{\rm obs}$ of observed unresolved objects in
three rings around NGC 1399. The error of $n_{\rm obs}$
comes from the poisson error of $N_{\rm obs}$ which is its square
root. The product of geometric ($\frac{n_{\rm obs}}{n_{\rm sel}}$) and
photometric (0.79) completeness, the total completeness $C_{\rm 0}$, is given in the last 
column.}
\begin{tabular}{ccccc}
\hline\small
Ring [$'$]& $N_{\rm obs}$&$n_{\rm obs}$ [arcmin$^{-2}$] & $C_{\rm 0}=\frac{n_{\rm obs}}{n_{\rm sel}}\cdot0.79$\\\hline\hline
2--8 & 13&0.069 $\pm$ 0.023 & 0.28 $\pm$ 0.08 \\
8--14 & 34&0.082 $\pm$ 0.014 & 0.33 $\pm$ 0.07 \\
14--20 & 19&0.030 $\pm$ 0.007 & 0.12 $\pm$ 0.03 \\\hline
2--20 & 66&0.053 $\pm$ 0.007 & 0.21 $\pm$ 0.04 \\
\normalsize
\end{tabular}
\end{table}To get $n_{\rm obs}$ for the three rings, we counted the number
$N_{\rm obs}$ of point sources within each one and divided this value
by the respective area: $n_{\rm obs}=\frac{N_{\rm obs}}{Area}$. These
numbers, together with the magnitude independent geometric
completeness $\frac{n_{\rm obs}}{n_{\rm sel}}$, are given in Table
\ref{nobserv}. Multiplying photometric and magnitude independent geometric
completeness yields the total magnitude independent completeness
$C_{\rm 0}$ of our survey, ranging between 12 and 33\%. The exact
values for the three rings are summarized in Table~\ref{nobserv}.\\
Up to now, the magnitude dependence of the geometric completeness has
not been considered; bright selected objects get higher priorities in
the mask creation than faint ones, and the faint sources more likely have too low S/N. To take this into account, we
subdivide the magnitude regime $19<V<21$ in two bins of 1 mag
width. For $19<V<20$ mag the geometric completeness is 1.18$\cdot
C_{\rm 0}$, for $20<V<21$ mag it is 0.92$\cdot C_{\rm 0}$.\\
\begin{figure}[h!] 
\begin{center}
\psfig{figure=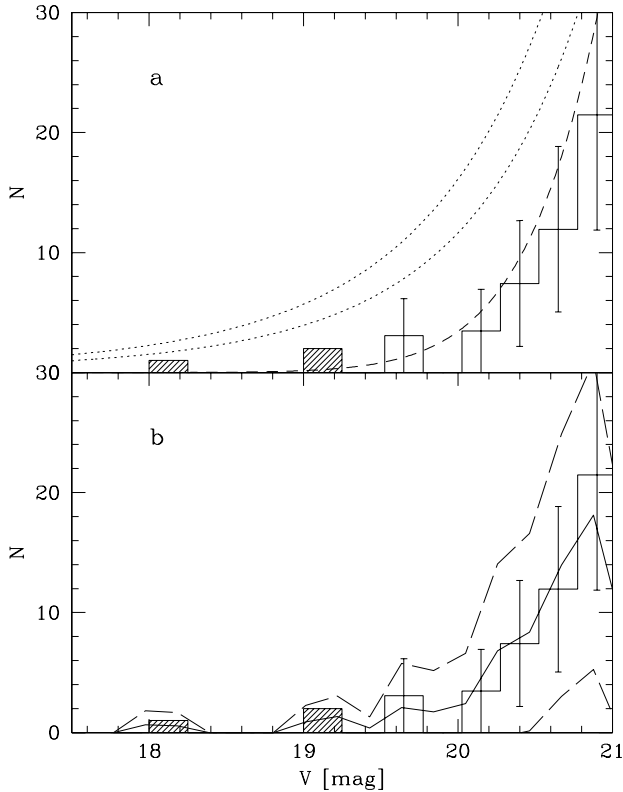,width=8.6cm}
\end{center}
\vspace{-0.2cm}
\caption[]{\label{lfgausst5}{\bf a)}: {\it Solid histogram}: Incompleteness
corrected LF of our GC candidates. Error bars are indicated. {\it 
    Shaded histogram}: LF of the UCOs within 20$'$ of NGC 1399. $Dotted$ $lines$: 
$t_5$-function with $V_{\rm to}$=23.9 mag,
$\sigma$=1.1 mag (upper line) and $\sigma$=1.0 mag (lower line). {\it
Short dashed line}: Gaussian with $V_{\rm to}$=23.9 mag and
$\sigma$=1.2 mag. Adopted total number of GCs:
8100 (see text). {\bf b)}: {\it Solid line}: 
Kernel estimator for the
incompleteness corrected LF of our GC candidates, using an
Epanechnikov kernel of width 0.125 mag. The error range is indicated
by the long dashed curves. Histograms as in panel a).}
\end{figure}
\subsubsection{Observed number of Globular Clusters}
\label{resexpnum}
As result of our radial velocity measurements we found 6 GC candidates
inside ring 1, 5 in ring 2, and 1 in ring 3. If they are all GCs, the
expected number of observed GCs should be close to these values for
each ring.\\
In the last three subsections, the {\it existing} number of GCs in
each of the three rings around NGC 1399, the photometric and the
geometric completeness were calculated. The number of {\it expected}
GCs is the number of existing ones multiplied by 0.8$\%$ (the bright
end of the LF) and the total completeness. As only one of the 12 GC
candidates is brighter than $V=20$ mag, 0.92$\cdot C_{\rm 0}$ is
adopted as the completeness
 (see Sect.~\ref{geometricincompleteness}). The results of these
calculations are given in Table \ref{nexpected}.\\
\begin{table*}
\begin{center}
\begin{tabular}{ccccc}
\hline Ring [$'$]&$N_{\rm exist}\cdot 0.008$&$C_{\rm
0}$$\cdot0.92$&$N_{\rm exp}$&$N_{\rm obs}$\\\hline\hline 2--8&30 $\pm$
6&0.253 $\pm$ 0.082&7.6 $\pm$ 2.8 &6\\ 8--14&19 $\pm$ 4&0.305 $\pm$
0.074&5.8 $\pm$ 2.4 &5\\ 14--20&16 $\pm$ 4&0.112 $\pm$ 0.040&1.8 $\pm$
1.4 &1\\\hline &&&15.2 $\pm$ 3.9&12\\
\end{tabular}
\end{center}
\renewcommand{\baselinestretch}{1.00} \small\normalsize
\caption[]{\label{nexpected}Number $N_{\rm exp}$ of expected GCs
compared to the number $N_{\rm obs}$ of observed GC candidates. The
errors of $N_{\rm exist}\cdot 0.008$ and $N_{\rm exp}$ are given by
their square root.}
\end{table*}
As one can see, the expected number of GCs matches well the number of
GC candidates in all three rings. $N_{\rm obs}$ is always
lower than $N_{\rm exp}$, but the error ranges overlap. In other words, the radial
distribution of our GC candidates agrees within the errors with the
distribution of the whole GCS. Summing the values up, the total
number of expected globulars is between 11 and 19 assuming a Gaussian
LF with $V_{\rm to}$=23.9 mag and $\sigma$=1.2, when we observe
12. This is a good agreement, too. It implies that the bright tail of the
GCLF is well described by a Gaussian and that the great majority of GC
candidates really are GCs.\\
There are of course other representation for the GCLF. Kohle et al. (\cite{Kohle96}) fit a $t_5$-function to their data and find that
the best value for $\sigma$ is 1.1 $\pm$ 0.1 mag, with the same
turnover than for the Gaussian. Hilker et al. (\cite{Hilker99}) made estimates of the
number of existing GCs in the bright manitude regime where they find 2
of the new UCOs ($V$=18.0 and $V$=19.1 mag, respectively). They adopt
both a Gaussian and a $t_5$-function and find that GCs as bright as
the brightest UCO can statistically only exist if the bright LF wings
are described by a $t_5$-function. The fraction of GCs brighter than
$V$=21 mag, when adopting the $t_5$-function as in Kohle et al.
becomes 2.3\% (1.7\% for $\sigma$=1.0 mag). This
is almost three times higher than for
the Gaussian. The expected number of observed GCs in our survey would become 44 $\pm$ 8, by
far higher than our
result, even when including the UCOs as possible GCs.\\
In Fig.~\ref{lfgausst5}a, the Gaussian ($\sigma$=1.2 mag) and the
$t_5$ functions ($\sigma$=1.1 and $\sigma$=1.0 mag) are plotted
together with the two luminosity functions. For the total number of
GCs we adopt 8100, which is the sum of the values calculated for the
three rings around NGC 1399 (see Sect.~\ref{totalnum}). The LF of our GC candidates is multiplied by the
total completeness as determined in
Sect.~\ref{geometricincompleteness}. Apparently, the LF is fit better
by the adopted Gaussian than by the $t_5$ functions, which
significantly overestimate the number of bright GCs. The existence of
GCs as bright as $V=18$ mag is therefore very unlikely.\\
\subsection{A gap between the UCOs and our GC candidates?}
\label{numcomp}
From Fig.~\ref{lfgausst5} one can see that the binning
independent representation of the joint LF of GCs and UCOs is very well fit by the Gaussian. In the magnitude regime between $19.2$ and $20.2$ mag there is no
evidence for a pronounced difference between the Gaussian and the
data. We can therefore not confirm a significant gap in magnitude
space between our newly found GC candidates and the 2 UCOs. There seems to be a smooth transition between both
populations. This is consistent with the faint UCOs and the GC candidates
belonging to the same group of objects, namely the
brightest GCs of NGC 1399. However, the smooth transition observed by us is only a
necessary and not a sufficient condition to link UCOs and GCs.
Our data are consistent with a
slight enhancement of number counts between $V=19$ and $V=20$ mag due
to the presence of a small number of stripped nuclei, too (for further
discussion, cf. Sect.~\ref{discussion}).\\
\subsection{Line indices and metallicities}
\subsubsection{Line indices}
\begin{figure}[h!]
\psfig{figure=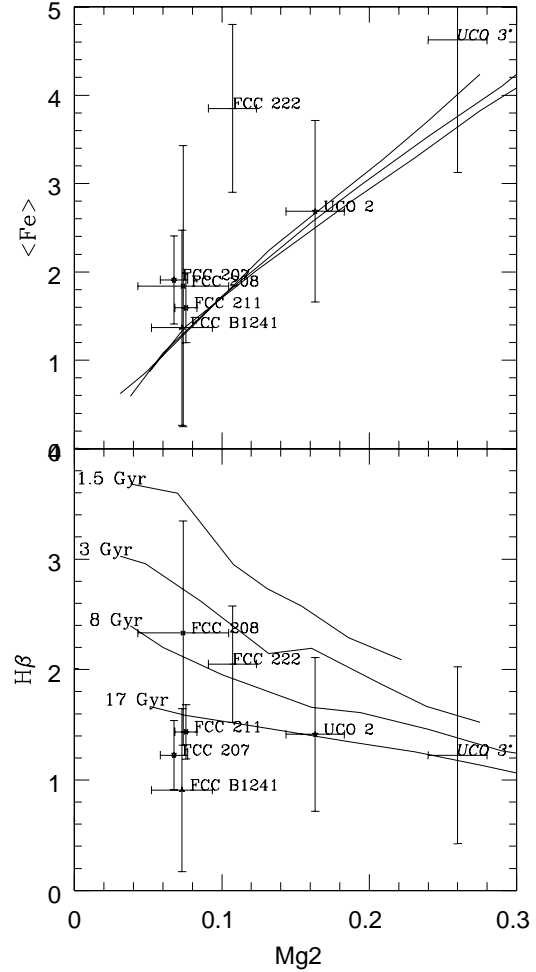,width=8.6cm}
\caption[]{\label{Mg2-Hb-Fe}{\it Upper panel}: Equivalent width of
H$\beta$ plotted vs. the line index Mg2. Evolutionary tracks for 1.5,
3, 8 and 17 Gyr as taken from Worthey (\cite{Worthe94}) are plotted as
reference. Their metallicity range is $-$2 ($-$1.7 for 1.5 and 3 Gyr)
$< [\frac{Fe}{H}] <$ 0.5 dex. {\it Lower panel}: $<Fe>$ plotted
vs. Mg2. Evolutionary tracks for 3, 8 and 17 Gyr (Worthey \cite{Worthe94}) are
overplotted for reference.\\$^*$ \small Values for UCO~3 from Hilker et
al. (\cite{Worthe94})\normalsize}
\end{figure}

One of the UCOs (UCO~2) was included in our survey. In this section
 its line indices are compared with the observed Fornax dE,Ns and the
 brightest GC candidate (object FCOS~1-021). The S/N of the other GC
candidates is too low (4 - 6) to reliably measure line indices. 
In order to estimate metallicities and ages, we measured 8 line indices
as defined by Faber and Burstein (\cite{Faber73})
and Brodie and Hanes (\cite{Brodie86}).\\
Brodie \& Huchra (\cite{Brodie90}) determine a linear metallicity calibration for
7 of the 8 indices we used (not for Fe53), based on the Zinn \& West
(\cite{Zinn84}) scale for Galactic Globular Cluster metallicities:
[$\frac{Fe}{H}$]=$a\cdot I+b$. Here, $I$ is the line index, $a$ and $b$ are
the coefficients of the linear calibration.\\
To measure the line indices and their error for each of the confirmed
Fornax members we followed closely the reduction procedure of Brodie
\& Huchra (1990). The object spectra were flux calibrated and all
bandpasses where shifted according to the objects' radial
velocity. 
The statistical error was determined from the
original, not flux calibrated spectra.\\
In Fig.~\ref{Mg2-Hb-Fe}, the equivalent widths of H$\beta$ (lower
panel) and $<Fe>$ (=(Fe52+Fe53)/2) (upper panel) are plotted
vs. Mg2. The values for UCO~3 as
determined by Hilker et al. (\cite{Hilker99}) are plotted as well. The red cutoff of object FCOS~1-021's spectrum is at about 5200
\AA, therefore the measurement of Mg2 and $<Fe>$ is not possible for
this object. For comparison, evolutionary tracks for different ages,
taken from Worthey (\cite{Worthe94}) are overplotted. The metallicity range is
$-$2 $< [\frac{Fe}{H}] <$ 0.5 dex for 17 and 8 Gyr and $-$1.7 $<
[\frac{Fe}{H}] <$ 0.5 dex for 3 and 1.5 Gyr. \\
Of the 6 plotted objects, only FCC~207 and FCC~211 (both dE,Ns), have
sufficiently high S/N to make a reliable age estimation. They appear
to be quite old (age $\geq$ 12 Gyr) with metallicities between $-$1.0
and $-$1.5 dex according to Worthey's values. One can see that UCO~2
shows a high metallicity in comparison to the other objects (about
$-$0.5 in dex), but appears to be old as well, although error bars are
large. The metallicity is calculated more accurately in the next
section. UCO~3 (the brightest UCO, value from Hilker et al. \cite{Hilker99}) appears
very old, too. It is even more metal rich than UCO~2.\\
In the $<Fe>$ vs. Mg2 plot, five of our six objects and UCO~3 fall nicely into
Worthey's evolutionary tracks, only FCC~222 apparently has a very high
Fe abundance compared to the other objects. \\
\subsubsection{Metallicities derived from line indices}
We calculated the metallicity for each index (except Fe53) using the
coefficients defined by Brodie and Huchra (\cite{Brodie90}) and then determined
the weighted average $[\frac{Fe}{H}]_{\rm W}$ of the six different
values.\\
The results for the seven Fornax members are given in Table
\ref{resmet}. Only for UCO~2, FCC~207 and FCC~211 and FCC~B1241 the
weighted means have errors smaller than 0.5 dex. The metallicities
of FCC~207 and FCC~211 lie well in the range of metallicities derived
spectroscopically for bright Fornax dEs. Examples are Held \& Mould
(1994) ($-$0.75 to $-$1.45), Brodie \& Huchra (\cite{Brodie91}) ($-$1.11 $\pm$
0.22) or the most recent study from Rakos et al. (\cite{Rakos01}) ($-$0.4 to
$-$1.6), the latter one based on Str\"omgren photometry. Held \& Mould
get $[\frac{Fe}{H}]$=$-$1.19 $\pm$ 0.05 for FCC~207, Rakos et
al. get $-$1.50 for both FCC~207 and FCC~211 and obtain
$[\frac{Fe}{H}]$=$-$0.9 for FCC~222. Thus, our values are consistent
with the findings of other groups.\\
\begin{table}
\begin{center}
\begin{tabular}{cc}
\hline
Name&$[\frac{Fe}{H}]_{\rm W}$ [dex]\\\hline\hline
UCO~2 & $-$0.57 $\pm$ 0.28\\
FCC~B1241   & $-$1.41 $\pm$ 0.37\\
FCC~208  & $-$1.85 $\pm$ 0.74\\
FCC~207  & $-$1.34 $\pm$ 0.41\\
FCC~211  & $-$1.39 $\pm$ 0.24\\
FCOS~1-021   & $-$0.17 $\pm$ 0.74\\
FCC~222  & $-$0.74 $\pm$ 0.68\\\hline
\end{tabular}
\end{center}
\renewcommand{\baselinestretch}{1.00}
\small\normalsize
\caption[]{\label{resmet}Results for the weighted mean $[\frac{Fe}{H}]_{\rm W}$ derived from 6 different line indices (see text).}
\renewcommand{\baselinestretch}{1.00}
\small\normalsize
\end{table}
UCO~2 is metal rich compared to average dE,Ns. For NGC 1399, the average GC metallicity is about
$-$1.0 $\pm$ 0.2 dex (Brodie \& Huchra \cite{Brodie91}, Kissler-Patig et
al. \cite{Kissle98}), but the metallicity distribution is bimodal with peaks at
about $-$1.5 and $-$0.5 dex (Kissler-Patig et al. \cite{Kissle98}). Thus, UCO~2
is more metal rich than the mean, but could belong to the metal rich
GC population. Hilker et al. (\cite{Hilker99}) find that UCO~3 is with
$[\frac{Fe}{H}]$ $\simeq$ 0 significantly metal richer than the dE,Ns,
while UCO~4 shows similar line index values like the
dE,Ns, however with very large errors.\\
The most compact known dE M~32 is known to have a metallicity close to
the solar one (e.g. del Burgo et al. \cite{delBur01}). Thus, the relatively high
metallicities of UCO~2 and 3 may not be too surprising, if it is an intrinsic property of these very compact
objects. The nucleus dominated spectrum of FCC~222 shows a high
metallicity (with large error) as well, so UCO~2 and 3 could be nuclei
of entirely stripped dE,Ns like FCC~222, too.\\
\section{Discussion}
\label{discussion}
What can we learn about the nature of the UCOs from our data?\\
The statistical arguments presented in Sect.~\ref{numgc} suggest that all our GC
candidates can be accounted for by GCs and that there is no significant
gap between our GC candidates and the fainter UCOs: the expected number of GCs detected in our survey (15 $\pm$ 4) is very
close to the number of our GC candidates (12) when assuming a Gaussian
LF for the GCS around NGC 1399. Adopting a LF with more extended wings
like a $t_5$-function overestimates the number of bright globulars by far (44 $\pm$ 8). We therefore conclude that the luminosity function at
the bright end is significantly better fit by a Gaussian than a
$t_5$. This makes the existence of bright GCs with magnitudes like the
{\it brightest} UCO~($M_{V}=-13.2$ mag) very unlikely.\\
The {\it fainter} UCOs ($M_{V}\simeq-12$ mag) do not appear as special
as they were thought to be, considering the shape of the LF as shown
in Fig.~\ref{lfgausst5}c. It is remarkable in this context that
Minniti et al. (\cite{Minnit98}) already found two GC candidates with $V=19.6$
and $V=19.8$ mag ($M_V$=$-$11.7 and $-$11.5 mag) about 5$'$ away from
NGC 1399 in a region not covered by us in this survey. This supports
our conclusions that there exists no significant gap between GCs and
faint UCOs and that our incompleteness corrections are in the right
order. Including the 2 UCOs outside 20$'$ we have confirmed Fornax
members at $V$ magnitudes of 19.1, 19.2, 19.3, 19.5, 19.6, 19.7, 19.8, 20.2
and fainter. The distinction between both groups of objects
based only on their different brightnesses might not be appropiate. The FCSS
(Drinkwater \cite{Drinkw00a}) was not deep enough to detect the extention of
the UCOs to fainter magnitudes. Therefore the detected UCOs seemed to be
a separate group of objects. Now we can state that the GCS of NGC 1399 is so rich that it
probably contains GCs as bright as the four fainter UCOs. However, this does
not exclude that some of them are isolated nuclei of dissolved dE,Ns.\\
  To link the UCOs to GCs, they not only have to share the same magnitude
  space, but need to have a mean radial velocity and spatial distribution consistent with the GCS of NGC 1399. Both
  properties can only be determined very unprecisely for a sample of five
objects. Nevertheless we remark
that the UCOs' mean radial velocity of 1530 $\pm$
  110 km/s (Drinkwater et al. \cite{Drinkw00a}) is consistent with that of
  the whole GCS (1447 $\pm$ 16 km/s, Richtler et al. 2001 private
  communication). Drinkwater et al. (\cite{Drinkw00a}) find that the UCOs
  are significantly more concentrated towards the cluster center than the
  Fornax dE,Ns, but have a shallower distribution than the GCS as derived
  from previous surveys (Grillmair et al. \cite{Grillm94}, Forbes et al. \cite{Forbes98}). However, the
  latest results by Dirsch et al. (\cite{Dirsch01}) show that the GCS of NGC
  1399 does extend to well beyond 20$'$ projected distance from the
  galaxy's center. The distribution of our 12 GC candidates is consistent
  with that (cf. Section~\ref{resexpnum}). Therefore even the furthest UCO
  at 28.3$'$ projected distance can still be a member of the GCS.\\
In the FCSS, Drinkwater et al. (\cite{Drinkw00a}) found that for all the UCOs,
cross correlation yielded higher confidence levels when using K-type
stars as templates than younger F-type stars. We can confirm this
result in so far as that cross correlation with HD~54810, a K star,
gives higher $R$
values than for HD~22879, an F star. For the nucleated dEs included in our survey it is vice
versa. Thus, the integrated stellar type of UCO~2 appears to be more similar
to those of GCs than to those of dE,Ns.\\
If nuclei of entirely stripped dwarf galaxies exist they would be
expected to mix up with the bright GCs, because on average they are
more than 2 mag brighter than GCs (cf. Introduction). The results
obtained by us do not speak against the existence of these "naked"
dwarf elliptical nuclei. Depending on the frequency of stripping, a
certain fraction of the UCOs and the GC candidates observed by us,
could be accounted for by stripped nuclei. We could be observing the
sum of the bright tail of the GCLF and the nucleus-LF. Without
thorough theoretical treatment of the frequency of entire stripping,
this is very hard to quantify, although the nuclei are certainly less
frequent than GCs. The number of dE,Ns in the central Fornax region is
only in the order of a few dozen (as well in Virgo, see Lotz et al. \cite{Lotz01}). The timescale for a total
stripping is in the order of several Gyr (Bekki et al. \cite{Bekki01}). So the
number of stripped nuclei should be much smaller than the number of
GCs, but due to their higher average brightness might become
significant at bright magnitudes. A possible effect of the "naked
nuclei" would be that the number counts at the very bright end
($V\simeq 19.5$ mag) are enhanced with respect to the extrapolation of
the GCLF from the magnitude range between $V=20$ and $V=21$ mag to
fainter magnitudes. With our limited statistical sample this cannot be
traced. We have only probed about 25\% of the area within 20$'$ from
NGC 1399. A much better coverage of the inner 20$'$ and extension to a
distance of 30$'$ (where the remaining two UCOs at $V\simeq 19.4$ mag
are found) is needed to make more definite statements. Although our
results together with Minniti et al.'s (\cite{Minnit98}) detections are
indicative for an extension of NGC 1399's GCS to about $M_{V}=-12$
mag, they have to be confirmed on a more profound statistical
basis. Observing time to substantially increase the area coverage of
  our survey has been approved.\\
The brightest UCO (UCO~3) clearly stands out from the others. It is
more than 1 mag brighter than the rest and based on our results it cannot
be explained by the GCS. UCO~3 thus probably is galaxian, either the
compact remnant of a once extended dE,N ("naked nucleus") or a cdE
which was already "born" as compact as it is now, or a non nucleated
dE changed by some tidal process to a more compact shape.\\
From the metallicity and morphology of the UCOs no conclusive statement can
be made:
our metallicity value for UCO~2 is quite high (about $-$0.55 in dex),
implying a metallicity higher than the majority of Fornax
dE,Ns. Still, dE,Ns with similar metallicities have been found (Rakos
et al. \cite{Rakos01}). Of special interest is that M~32 -- the most similar dwarf
galaxy to the UCOs -- has a very high metallicity, about solar
(Burgos et al. 2001). As well, UCO~2 could belong to the metal rich GC
population of NGC 1399 (Kissler-Patig et al. \cite{Kissle98}). No discriminating
statement can be made from metallicity, and due to the still too low
S/N of the spectrum, a reliable age determination is not possible.\\
In our images, UCO~3 is the only one which is not classified as a
stellar-like object according to the selection criteria given in
Sect.~\ref{Selcand}. Its FWHM is 2.3$''$, when the seeing was
1.85$''$. This corresponds to a diameter of about 170 pc, by far
larger than any GC found so far. But, the other UCOs are all
unresolved, they have effective radii of about 15 pc (HST-STIS data
published in Drinkwater et
al. \cite{Drinkw01b}). None of the UCOs shows low surface brightness features in
the outer part away from the central peak, although we can detect features
down to $\simeq$ 26.5 mag/arcsec$^2$ in $V$ (Hilker et al. \cite{Hilker02}, in prep.). If the UCOs are remnants of
dE,Ns, the stripping has been extremely efficient.\\
\section{Summary and conclusions}
\label{summary}
In this paper, we presented a spectroscopic survey on compact objects
in the central region of the Fornax cluster. The aim was to survey the
magnitude regime framed by the newly discovered ultra compact objects
(UCOs) in Fornax (Drinkwater et al. \cite{Drinkw00a}, Hilker et al. \cite{Hilker99}) and the
brightest GCs around NGC 1399. In the FCSS, performed by Drinkwater et
al., the UCOs are at the faint magnitude limit of their survey. We
wanted to know whether these detections constitute the bright tail of
a continous luminosity distribution with a smooth transition into the
GC regime or whether they are a separate population.\\
The velocity measurements (cf. Sect.~\ref{measureradvel}) resulted in
the discovery of 12 GC candidates in the magnitude range between
$19.70<V<20.95$ mag ($-11.6<M_{V}<-10.35$), located all within 20$'$
from NGC 1399. For four of them, cluster membership was known
before. Their mean colour is $<V-I>$ = 1.11 $\pm$ 0.11 mag and their
mean radial velocity 1300 $\pm$ 109 km/s with $\sigma$=377 km/s.\\
In the subsequent analysis of the discoveries
(cf. Sect.~\ref{analysis}), the following results were obtained:\\
1. The expected number of observed GCs originating from NGC 1399's GCS
   is 15 $\pm$ 4, in good agreement with the 12 objects found. To
   calculate the expected number we assumed a Gaussian LF with $V_{\rm
   to}$ = 23.9 mag and $\sigma$ = 1.2 mag taken from Kohle et
   al. (\cite{Kohle96}), used the GC surface density from Dirsch et al. (\cite{Dirsch01})
   and took into account photometric and geometric incompleteness.\\
2. Assuming a more extended $t_5$ distribution for the LF as adopted
   by Kohle et al. (\cite{Kohle96}), the expected number of observed GCs rises
   to 44 $\pm$ 8, which is more than three $\sigma$ higher than the
   number of observed objects. This implies that the LF has no
   extended bright wings. Hilker et al. (\cite{Hilker99}) found that only by
   assuming such an extended LF the brightest UCO can be explained as
   a GC.\\
3. There is no significant gap in magnitude space between our GC
   candidates and the four fainter UCOs within 20$'$ of NGC 1399. The
   GCS of NGC 1399 appears to extent to $M_{V}\simeq -12$ mag.\\
4. The only UCO included in our survey is slightly better fit by early
   stellar types, in contrast to the dE,Ns.\\
5. The only UCO included in our survey has a relatively high
   metallicity compared to the dE,Ns, but is in the range of metal rich
   GCs or very compact dwarfs with almost solar metallicity like
   M~32.\\
Considering only point 5, we can neither rule out nor confirm that
the UCOs are bright GCs. From points 1 to
4 we conclude: the four UCOs fainter than $V=19.1$ mag
($M_{V}=-12.2$ mag) can be well explained by the bright tail of the
GCLF of NGC 1399. However, the apparent overlap of the two LFs is not
  sufficient to exclude the existence of stripped nuclei
of formerly extended dE,Ns mixing up with the bright GCs. The faint UCOs are probably no extremely
faint examples of cdEs, but are the brightest members of their object
class (GCs or stripped nuclei). UCO~3, however, is so bright and
large, that it probably is not a GC. It remains the most puzzling
object.\\\\

\acknowledgements We thank our referee Michael Drinkwater for his useful
comments which helped to improve the paper. SM acknowledges support by the
Heinrich-Hertz-Stiftung of the Ministerium f\"ur Bildung und
Wissenschaft des Landes Nordrhein-Westfalen. MH and LI acknowlegde
support through ``Proyecto FONDECYT 3980032'' and ``8970009'',
respectively. LI thanks {\it Proyecto Puente PUC}, CONICYT and
Pontificia  Universidad Cat\'olica de Chile for partial funding.\\

\newpage
\begin{appendix}
\section{Tables of all foreground stars and background galaxies}
Table~\ref{foreground} contains the parameters of all foreground stars and
Table~\ref{background} of all background galaxies for which a radial velocity
was determined.\\
The name given in the first column consists of the acronym
FCOS (Fornax Cluster Compact Survey) followed by a sequence number of the
field and the object reference number used in the mask creation process. An
asterix $^*$ means that the object's radial velocity could be determined
only by identification of emission lines, because cross
correlation yielded unreliable results. These were objects that show $one$ strong emission line on top of a faint
continuum. In all cases with stronger continuum and clearly identifiable
absorption lines present (H\&K), this emission line proved to be the
redshifted OII line at restframe 3727.3 \AA. Therefore, whenever the
continuum was faint, we looked for absorption features close to the emission
line and calculated their restframe wavelength assuming that the emission
line was OII. If the wavelenght were equal to important absorption
lines around 3727 {\AA} such as H8, H9 and H10, we accepted the assumption
that the emission line was OII and determined the corresponding
redshift. When no reliable identification was possible, no redshift
determination was done.\\
The second and third column are the right
ascension and declination for epoch 2000.\\
In the fourth column, the radial
velocity with its error is given. Both values were computed by
averaging the values given from FXCOR for each of the three templates used
for cross correlation. In case of emission line objects with very faint
continuum, the error was derived from the uncertainty of the emission
line position. For the four quasars discovered,
redshift instead of radial velocity is given.\\
In the fifth and sixth column, apparent magnitude $V$ and colour $(V-I)$ are
given. Both values are from Hilker et al. (\cite{Hilker02}, in prep.).\\
Table~\ref{background} contains an additional comments-column. Here, ELO
stands for ``Emission Line Object''. Quasars were identified by their
extremely broad emission line features.\\ 
\renewcommand{\baselinestretch}{0.9}
\begin{table*}
\caption{\label{foreground}Foreground stars.}
\begin{tabular}[l]{lrrrrr}
Name&$\alpha$ (2000.0)&$\delta$ (2000.0)&$v_{\rm rad} [km/s]$&$V$&$(V-I)$\\\hline\hline
  FCOS~6-052 & 3:34:55.50 &     $-$35:03:59.6 &     230 $\pm$     60 &      20.70 &     0.52\\            
  FCOS~6-042 & 3:34:57.65 &     $-$35:03:23.0 &     195 $\pm$     45 &     20.35 &     0.62\\            
  FCOS~6-050 & 3:35:01.50 &     $-$35:07:54.0 &     170 $\pm$     25 &     20.66 &     1.26\\            
  FCOS~6-043 & 3:35:03.10 &     $-$35:17:31.9 &     325 $\pm$     75 &     20.36 &    $-$0.13\\        
  FCOS~6-047 & 3:35:03.87 &     $-$35:14:01.7 &     180 $\pm$     25 &     20.59 &     1.07 \\            
  FCOS~6-041 & 3:35:05.55 &     $-$35:10:10.6 &     120 $\pm$     25 &     20.31 &     1.04 \\            
  FCOS~6-029 & 3:35:07.03 &     $-$35:06:16.8 &      45 $\pm$     25 &     19.46 &      0.70 \\            
  FCOS~6-048 & 3:35:07.30 &     $-$34:58:13.1 &     130 $\pm$     25 &     20.61 &     0.87 \\            
  FCOS~6-038 & 3:35:09.32 &     $-$35:18:47.1 &     175 $\pm$     35 &     20.27 &     0.55 \\            
  FCOS~6-030 & 3:35:13.04 &     $-$35:04:46.3 &     135 $\pm$     25 &     19.49 &     1.38 \\            
  FCOS~6-034 & 3:35:13.99 &     $-$35:05:02.6 &      80 $\pm$     15 &      19.80 &     1.11 \\            
  FCOS~6-027 & 3:35:17.25 &     $-$35:20:20.8 &     240 $\pm$     35 &     19.06 &     0.61 \\            
  FCOS~6-051 & 3:35:18.31 &     $-$34:58:21.7 &     160 $\pm$     35 &      20.70 &     0.55 \\            
  FCOS~6-037 & 3:35:19.47 &     $-$34:59:37.2 &     115 $\pm$     15 &     20.17 &     1.44 \\            
  FCOS~6-039 & 3:35:22.54 &     $-$35:07:03.6 &      75 $\pm$     15 &     20.27 &     0.69 \\            
  FCOS~6-033 & 3:35:22.66 &     $-$35:06:34.3 &      95 $\pm$     15 &     19.76 &     1.46 \\            
  FCOS~6-056 & 3:35:24.22 &     $-$35:16:50.8 &     135 $\pm$     25 &     20.88 &     1.12 \\            
  FCOS~6-066 & 3:35:25.08 &     $-$35:21:46.2 &     $-$35 $\pm$     15 &     16.45 &     1.14 \\            
  FCOS~6-028 & 3:35:26.75 &     $-$35:16:35.8 &      85 $\pm$     15 &      19.2 &     1.38 \\            
  FCOS~6-036 & 3:35:32.04 &     $-$35:16:01.0 &     110 $\pm$     25 &     19.96 &     0.53 \\            
  FCOS~2-050 & 3:37:42.65 &     $-$35:25:41.4 &     190 $\pm$     25 &     19.34 &     0.69 \\            
  FCOS~2-080 & 3:37:42.98 &     $-$35:42:14.5 &     140 $\pm$     45 &      20.70 &     0.59 \\            
  FCOS~2-105 & 3:37:43.04 &     $-$35:38:28.6 &     135 $\pm$     35 &     19.19 &     0.64 \\            
  FCOS~2-059 & 3:37:43.49 &     $-$35:25:57.7 &     190 $\pm$     35 &     19.71 &     1.47\\            
  FCOS~4-021 & 3:37:44.01 &     $-$35:15:09.3 &      95 $\pm$     25 &     19.81 &     1.04\\            
  FCOS~2-051 & 3:37:45.96 &     $-$35:27:23.4 &     315 $\pm$     35 &     19.42 &     0.86 \\            
  FCOS~4-037 & 3:37:46.43 &     $-$35:13:01.4 &      20 $\pm$     35 &      20.40 &     1.23\\            
  FCOS~4-033 & 3:37:47.42 &     $-$35:17:18.1 &      80 $\pm$     25 &      20.30 &     1.26 \\            
  FCOS~2-054 & 3:37:49.39 &     $-$35:44:07.5 &     145 $\pm$     25 &     19.55 &     1.21 \\            
  FCOS~4-043 & 3:37:49.66 &     $-$35:19:43.0 &     165 $\pm$     70 &
  20.7 &     0.59\\
  FCOS~4-036 & 3:37:51.18 &     $-$35:12:42.3 &     $-$35 $\pm$     60 &     20.37 &      0.60\\            
  FCOS~4-026 & 3:37:51.66 &     $-$35:05:19.3 &      95 $\pm$     25 &     20.02 &     0.68\\            
  FCOS~2-057 & 3:37:52.02 &     $-$35:33:07.7 &     125 $\pm$     15 &     19.65 &     0.99\\            
  FCOS~4-029 & 3:37:52.49 &     $-$35:07:34.3 &      70 $\pm$     45 &     20.06 &     0.64\\            
  FCOS~4-031 & 3:37:53.43 &     $-$35:22:00.7 &     270 $\pm$     50 &     20.27 &     0.72\\            
  FCOS~4-027 & 3:37:54.57 &     $-$35:18:25.4 &     140 $\pm$     35 &     20.02 &     1.18\\            
  FCOS~4-020 & 3:37:55.11 &     $-$35:16:57.7 &     145 $\pm$     85 &     19.27 &     0.07\\        
  FCOS~4-056 & 3:38:00.85 &     $-$35:00:10.1 &     100 $\pm$     45 &     16.48 &     0.93\\            
  FCOS~4-022 & 3:38:01.58 &     $-$35:02:23.3 &      20 $\pm$     25 &     19.82 &     1.45\\            
  FCOS~2-070 & 3:38:05.23 &     $-$35:35:31.8 &      20 $\pm$    110 &     20.21 &     1.45\\            
  FCOS~2-058 & 3:38:08.05 &     $-$35:33:28.9 &      $-$5 $\pm$     15 &     19.66 &     1.11\\            
  FCOS~4-041 & 3:38:08.40 &     $-$35:20:46.2 &      70 $\pm$     15 &     20.54 &      0.60\\            
  FCOS~4-028 & 3:38:09.46 &     $-$35:08:35.5 &      20 $\pm$     25 &     20.06 &     1.23\\            
  FCOS~4-035 & 3:38:10.01 &     $-$35:04:33.5 &      80 $\pm$     45 &     20.35 &     0.59\\            
  FCOS~2-079 & 3:38:12.93 &     $-$35:41:44.2 &     205 $\pm$     35 &     20.69 &     0.63\\            
  FCOS~2-060 & 3:38:15.50 &     $-$35:43:28.2 &     125 $\pm$     25 &     19.75 &     0.66\\            
  FCOS~1-035 & 3:38:38.45 &     $-$35:34:51.4 &      30 $\pm$     60 &     19.65 &     0.78\\            
  FCOS~1-028 & 3:38:40.67 &     $-$35:36:52.3 &     285 $\pm$     30 &     19.86 &     1.17\\            
  FCOS~1-023 & 3:38:42.69 &     $-$35:43:37.7 &     160 $\pm$     45 &     20.37 &     0.79\\            
  FCOS~1-033 & 3:38:43.67 &     $-$35:35:05.0 &     200 $\pm$     35 &     19.76 &     0.87\\            
  FCOS~1-061 & 3:38:45.81 &     $-$35:25:12.6 &     170 $\pm$     45 &     20.89 &     0.83\\            
  FCOS~1-047 & 3:38:45.91 &     $-$35:32:40.6 &      60 $\pm$     45 &     20.31 &     0.75\\            
  FCOS~1-024 & 3:38:46.11 &     $-$35:42:11.4 &      70 $\pm$     35 &     19.75 &     0.85\\            
  FCOS~1-073 & 3:38:48.85 &     $-$35:40:16.4 &     115 $\pm$     15 &     19.09 &     1.33\\            
  FCOS~1-049 & 3:38:49.43 &     $-$35:30:01.8 &      10 $\pm$     50 &     20.84 &     0.73\\            
  FCOS~1-057 & 3:38:51.15 &     $-$35:27:41.2 &     175 $\pm$     50 &     20.95 &     0.72\\        
  FCOS~1-025 & 3:38:59.26 &     $-$35:41:14.4 &     140 $\pm$     50 &     20.69 &     1.31\\            
  FCOS~1-043 & 3:39:02.39 &     $-$35:33:53.4 &     175 $\pm$     25 &     20.31 &     1.06\\            
  FCOS~1-041 & 3:39:06.29 &     $-$35:34:11.4 &     100 $\pm$     15 &     19.76 &     1.17\\            
  FCOS~1-030 & 3:39:07.31 &     $-$35:36:02.2 &     255 $\pm$     45 &     20.24 &     0.72\\            
  FCOS~1-053 & 3:39:10.81 &     $-$35:29:09.7 &     120 $\pm$     35 &     20.48 &     1.44\\            
  FCOS~1-066 & 3:39:12.55 &     $-$35:21:22.7 &      60 $\pm$     35 &
  19.63 &  1.19\\\hline\hline
\end{tabular}
\end{table*}

\begin{table*}
\caption[]{\label{background}Background galaxies}
\begin{tabular}[l]{lrrrrrl}
Name&$\alpha$ (2000.0)&$\delta$ (2000.0)&$v_{\rm rad} [km/s]$&$V$&$(V-I)$&Comment\\\hline\hline
FCOS~6-023$^*$ & 3:34:57.34 & $-$35:12:24.0 & 16830 $\pm$ 85 & 18.97 & 0.72 &ELO\\
FCOS~6-032$^*$ & 3:34:59.92 & $-$35:11:41.6 & 57820 $\pm$ 130 & 19.71 & 1.17 &ELO\\
FCOS~6-057$^*$ & 3:35:00.09 & $-$35:09:25.3 & z=2.17 & 20.95 & 0.70 &Quasar\\
FCOS~6-005$^*$ & 3:35:03.08 & $-$35:20:53.4 & 49220 $\pm$ 105 & 19.57 & 0.85 &\\
FCOS~6-006$^*$ & 3:35:03.29 & $-$35:00:33.8 & 94500 $\pm$ 175 & 19.69 & 0.76 &\\
FCOS~6-001 & 3:35:06.61 & $-$35:10:53.5 & 61840 $\pm$ 50 & 19.24 & 1.41 &\\
FCOS~6-045$^*$ & 3:35:26.93 & $-$34:59:51.0 & 100900 $\pm$ 85 & 20.52 & 1.30 &ELO\\
FCOS~6-031 & 3:35:31.62 & $-$35:02:49.9 & 56950 $\pm$ 50 & 19.66 & 1.34 &\\
FCOS~4-040$^*$ & 3:37:39.46 & $-$35:01:34.8 & 75515 $\pm$ 85 & 20.45 & 1.48 &\\
FCOS~2-055$^*$ & 3:37:42.69 & $-$35:32:29.2 & 71045 $\pm$ 140 & 19.58 & 1.13 &ELO\\
FCOS~4-007$^*$ & 3:37:43.02 & $-$35:13:56.1 & 82650 $\pm$ 155 & 20.27 & 1.44& \\
FCOS~4-023$^*$ & 3:37:43.30 & $-$35:11:02.1  & z=2.29  & 19.87 & 0.53 &Quasar\\
FCOS~2-052 & 3:37:44.46 & $-$35:36:18.0 & 48195 $\pm$ 45 & 19.43 & 1.33 & \\
FCOS~2-084$^*$ & 3:37:44.59 & $-$35:34:50.9 & 41635 $\pm$ 85 & 20.79 & 1.08
&ELO\\
FCOS~4-001$^*$ & 3:37:44.62 & $-$35:18:32.6 & 84230 $\pm$ 175 & 19.63 & 1.09 &ELO\\
FCOS~4-006$^*$ & 3:37:45.37 & $-$35:00:22.2 & 49950 $\pm$ 140 & 20.12 & 0.91 &ELO\\
FCOS~4-016$^*$ & 3:37:45.63 & $-$35:03:35.4 & 76910 $\pm$ 210 & 21.04 & 0.83 &ELO\\
FCOS~2-001$^*$ & 3:37:46.47 & $-$35:24:11.4 & 64460 $\pm$ 105 & 19.17 & 1.40 &ELO\\
FCOS~2-065$^*$ & 3:37:46.65 & $-$35:27:50.0 & 33960 $\pm$ 120 & 20.02 & 1.42 &\\
FCOS~2-071$^*$ & 3:37:48.12 & $-$35:33:57.8 & 41755 $\pm$ 120 & 20.24 & 0.86 &ELO\\
FCOS~2-031$^*$ & 3:37:48.21 & $-$35:29:12.1 & 64800 $\pm$ 140 & 20.50 & 0.91& ELO\\
FCOS~4-025$^*$ & 3:37:48.69 & $-$35:04:23.1 & z=2.57 & 19.97 & 0.92 &Quasar\\
FCOS~2-085$^*$ & 3:37:50.94 & $-$35:39:47.2 & 95380 $\pm$ 175 & 20.80 & 1.05 &\\
FCOS~4-044$^*$ & 3:37:52.48 & $-$35:13:13.6 & 87600 $\pm$ 140 & 20.72 & 1.00 &\\
FCOS~4-024 & 3:37:55.60 & $-$35:09:36.1 & 50165 $\pm$ 50 & 19.93 & 1.35 &\\
FCOS~4-008$^*$ & 3:37:55.70 & $-$35:06:53.0 & 92200 $\pm$ 120 & 20.31 & 1.24 &\\
FCOS~2-008 & 3:37:59.86 & $-$35:39:04.4 & 61900 $\pm$ 80 & 19.68 & 1.40 &\\
FCOS~4-010$^*$ & 3:38:02.74 & $-$35:01:19.2 & 33350 $\pm$ 140 & 20.45 & 0.58 &\\
FCOS~4-032$^*$ & 3:38:02.79 & $-$35:08:55.4 & 73095 $\pm$ 160 & 20.30 & 1.37 &ELO\\
FCOS~2-066$^*$ & 3:38:03.76 & $-$35:36:07.9 & 48325 $\pm$ 85 & 20.07 & 0.92 &ELO\\
FCOS~4-030$^*$ & 3:38:12.67 & $-$35:16:20.3 & z=2.53 & 20.13 & 0.62 &Quasar\\
FCOS~4-003$^*$ & 3:38:21.76 & $-$35:09:22.5 & 50260 $\pm$ 140 & 19.77 & 1.03 &\\
FCOS~1-011 & 3:38:31.12 & $-$35:39:13.4 & 33455 $\pm$ 85 & 18.08 & 1.14 &ELO\\
FCOS~1-007 & 3:38:42.45 & $-$35:30:50.2 & 33215 $\pm$ 50 & 20.10 & 1.24 &\\
FCOS~1-003$^*$ & 3:38:49.09 & $-$35:24:12.0 & 87600 $\pm$ 210 & 19.79 & 1.11 &ELO\\
FCOS~1-006$^*$ & 3:38:49.14 & $-$35:29:20.2 & 82330 $\pm$ 225 & 19.63 & 0.93 &ELO\\
FCOS~1-008$^*$ & 3:38:56.91 & $-$35:31:10.9 & 128801 $\pm$ 300 & 21.02 & 0.98 &ELO\\
FCOS~1-013$^*$ & 3:39:02.64 & $-$35:41:49.5 & 88492 $\pm$ 190 & 20.68 & 0.85 &ELO\\
FCOS~1-029$^*$ & 3:39:03.63 & $-$35:36:14.6 & 55305 $\pm$ 155 & 19.74 & 0.92 & ELO\\
FCOS~1-075 & 3:39:08.18 & $-$35:38:41.3 & 48075 $\pm$ 85 & 18.68 & 0.99 & ELO
\\\hline\hline
\end{tabular}
\end{table*}
\renewcommand{\baselinestretch}{1.0}
\end{appendix}
\end{document}